\documentclass{article}
\usepackage[utf8]{inputenc}
\usepackage{graphicx}
\usepackage{hyperref}
\usepackage{color}
\usepackage[numbers]{natbib}
\usepackage{moreverb}
\usepackage{xcolor}
\usepackage{comment}
\usepackage{framed}
\usepackage{ragged2e}
\usepackage{csquotes}
\usepackage{colortbl}
\usepackage{geometry}
\usepackage{wasysym}
\usepackage{longtable}
\usepackage{hyperref}
\usepackage{lipsum}
\usepackage{authblk}

\newcommand{\brendan}[1]{\textcolor{blue}{}}

\begin{document}

\author[1]{Kelly Blincoe}
\author[2]{Markus Luczak-Roesch}
\author[2]{Tim Miller}
\author[3]{Matthias Galster}

\affil[1]{The University of Auckland, New Zealand, k.blincoe@auckland.ac.nz}

\affil[2]{Victoria University of Wellington, New Zealand, markus.luczak-roesch@vuw.ac.nz, tim.miller@vuw.ac.nz}

\affil[3]{University of Canterbury, New Zealand, matthias.galster@canterbury.ac.nz}

\title{Human-centric Literature on Trust for SfTI Veracity Spearhead}
\date{}
\maketitle

\begin{abstract}
    This article summarizes the literature on trust of digital technologies from a human-centric perspective. We summarize literature on trust in face-to-face interactions from other fields, followed by a discussion of organizational trust, technology-mediated trust, trust of software products, trust of AI, and blockchain. This report was created for the Science for Technological Innovation Veracity Spearhead supported by New Zealand's National Science Challenges.
\end{abstract}

\section{Trust in General} 

\textit{Disclaimer: This report does not include a Māori perspective on trust. Work is in progress to combine this Western view with a Māori perspective.}
\vspace{1cm}

Trust has been described as facilitating cooperative behaviour~\cite{shneiderman2000designing}. Trust has been examined extensively in the fields of experimental psychology, philosophy, sociology, and political science~\cite{lewis1985trust}. In sociology, trust is considered to be multi-faceted with distinct cognitive, emotional, and behavioral dimensions~\cite{lewis1985trust}. The cognitive dimension says that trust is based on rational decisions, while the emotional dimension, which is also referred to as affect-based trust, says that emotional relationships between people form a basis for trust~\cite{mcallister1995affect, chowdhury2005role}. The behavioral component of trust is the action of doing something with uncertain outcomes while assuming that all people involved in the action will act with integrity~\cite{barber1983logic}. Trust is only needed when there is some level or risk or uncertainty involved, so trust also involves vulnerability~\cite{wang2005overview}. Research in psychology and philosophy also describe trust as having both rational and emotional aspects~\cite{trvcek2018brief}. 

\brendan{It'd be great/ importnat and consistent for us to  address trust from a cultural perspective and in particular Mātauranga Maori. I'm thinking this through the cultural anthropology lens. Getting along not only requires thinking about the differences between cultures, but also the differences between persons as kin, as strangers and other forms of relatedness. These differences are altered through culture, questions of trust, intimacy and the modes through which others are framed. I see Maui's writtne  apaper on some of this https://digitalcouncil.govt.nz/assets/Uploads/Maori-Perspectives-on-Trust-and-Automated-Decision-Making-13-Nov-2020-1.pdf}

\begin{figure}
    \centering
    \includegraphics{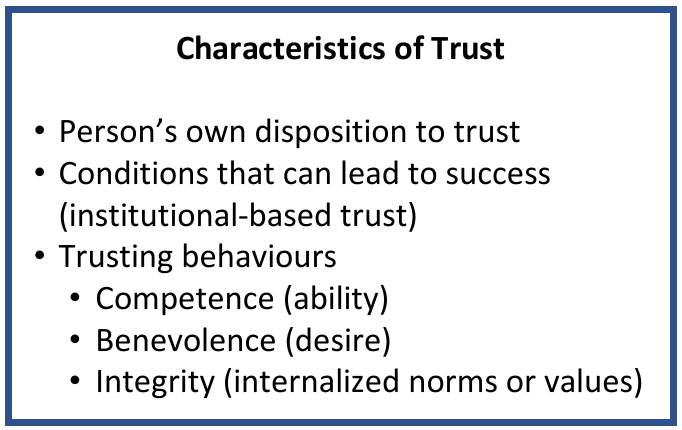}
    \caption{Characteristics of Trust~\cite{riegelsberger2005mechanics}}
    \label{fig:trust_general}
\end{figure}

Much research has considered how initial trust is formed. McKnight and Chervany identified a set of characteristics based on a review of literature across multiple disciplines (see Figure~\ref{fig:trust_general})~\cite{mcknight2001trust}. First, a person's own disposition to trust, or their willingness or tendency to depend on others, impacts how trust is formed. Second, there must be conditions in place that could lead to success, this is called institutional-based trust. Finally, there are three main trusting behaviours: competence, benevolence, and integrity. 
Competence is defined as the belief that the other party has the required skills, benevolence is the belief that the other party wants to do good, and integrity relates to the belief that the other party has good values or character. 

Studies have shown that people are influenced by a ``truth bias'', meaning that they are more likely to assess things as truth than lies, even when being deceived~\cite{mccornack1986deception, zuckerman1981verbal, zuckerman1984anchoring}.
People are not very good at identifying deception, which is defined as someone purposely misleading someone else~\cite{levine1999accuracy}. Research in psychology has found that people assume all information as truth initially and only later change their assessment if they find the information is false~\cite{gilbert1990unbelieving}. This is in line with the Truth-Default theory, which states that people tend to believe each other by default~\cite{levine2014truth}.

When trust is damaged, there are negative consequences~\cite{lewicki1996developing,robinson1996trust}. Researchers have studied how trust is repaired after it has been damaged~\cite{kim2009repair,tomlinson2009role}. Unlike initial trust, significant effort is often required to rebuild trust after a trust violation. Various factors impact trust repair, including the strength of the initial trust~\cite{lewicki2000trust}. A theory which is important for trust repair is attribution theory~\cite{weiner1985attributional}. Attribution theory considers the cause of the trust violation. It considers three main dimensions: locus of control (internal or external), controllability, and stability~\cite{tomlinson2009role}. Attribution theory states that the outcomes and reactions of trust violations will vary based on these dimensions. The theory also suggests that the outcomes and reactions are not permanent and that trust can be repaired following violations~\cite{bansal2015trust}.

\section{Organizational Trust}

While most research in the field of psychology has focused on interpersonal trust, organizational trust has been studied extensively in other domains like management and marketing~\cite{wang2005overview}. Organizational trust has been defined as ``the belief that the decision makers will produce outcomes favorable to the person’s interests without any influence by the person''~\cite{driscoll1978trust}. Management researchers argue that trust can improve business performance~\cite{wang2005overview,sako2006does}. Trust within an organization can result in improved productivity and satisfaction of employees~\cite{mayer1995integrative}. In the field of marketing, researchers have found that consumers' trust of a business is impacted by both the people within that business that they interact with and the business's management practices and policies~\cite{sirdeshmukh2002consumer}.

\section{Technology-Mediated Trust} 
\begin{figure}
    \centering
    \includegraphics[width=.5\columnwidth]{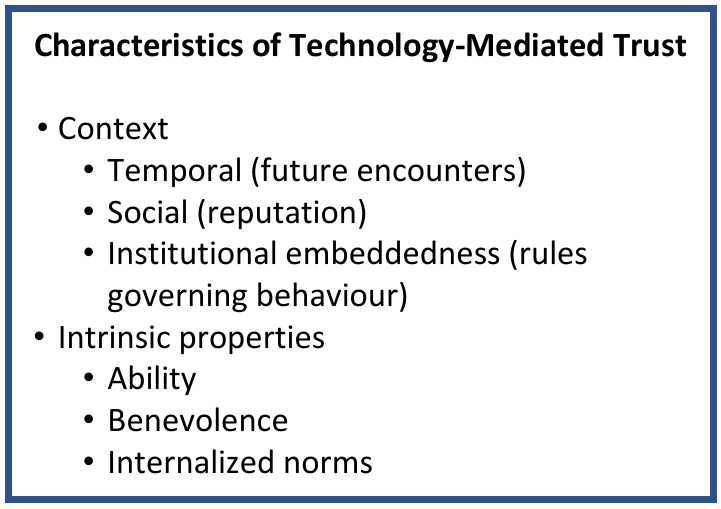}
    \caption{Characteristics of trust from Riegelsberger et al. framework for technology-mediated interactions~\cite{riegelsberger2005mechanics} }
    \label{fig:riegelsberger}
\end{figure}

Trust has primarily been studied from a perspective of human, face-to-face interactions~\cite{mcknight1996meanings}. When interactions occur through technology, signals of trust are different~\cite{shneiderman2000designing}.  
Riegelsberger et al. proposed a framework of trust in technology-mediated interactions (see Figure \ref{fig:riegelsberger}), which included both contextual and intrinsic properties of trust~\cite{riegelsberger2005mechanics}. The contextual properties included in the framework are temporal, social, and institutional embeddedness. Where temporal embeddedness considers likely future encounters since repeated interactions can both encourage trustworthy behaviour and provide signals to make decisions around trust. Social embeddedness considers the reputation of the person or organization who is being trusted, which can be discussed and shared across the trustors (those doing the trusting). Institutional embeddedness refers to the institutions that govern behaviour, such as judicial systems or organizations, since the rules imposed by these institutions will influence trust. Yet, there is an acknowledgement that new technology can disrupt trust formation, since new technology has the potential to transform the way in which people interact, which can lead to uncertainty and vulnerability until new norms are established~\cite{mcknight2000trust}. 

\brendan{There is also cultural embeddedness see summary doc https://www.redefinerswl.org/post/what-is-cultural-embeddedness-and-how-to-find-freedom-from-it that includes aspects of 'shared meaning'. Of course multiple langauges enables us to have multiple shared meanings, truths and trust frameworks. I wonder if one of our tasks is to explore a 'shared meaning' of veracity through multiple cultural lenses/ languages etc. Build a common form of trust.
Our diversity and willingness to work together is a perfect environment to experiment with this. A kind of living lab, not just a Veracity one.}

The intrinsic properties included in Riegelsberger et al.'s framework are ability, internalized norms, and benevolence, which are in line with the trusting behaviours of competence, integrity, and benevolence of McKnight and Chervany described above~\cite{mcknight2001trust}. Here, ability refers to the capabilities and characteristics of the person or organization who is being trusted that will enable them to fulfill the promised outcomes. Internalized norms includes attributes such as honesty, credibility, reliability, dependability, openness, and good will. Benevolence represents the enjoyment obtained by person or organization who is being trusted when good outcomes are experienced by the person doing the trusting.

While this framework was created to conceptualise trust in technology-mediated interactions, the elements of trust are still focused on the people or organizations involved in the trust relationship. Shneiderman also recognized this in his definition of trust: ``If users rely on a computer and it fails, they may get frustrated or vent their anger by smashing a keyboard, but there is no relationship of trust with a computer. If users depend on a network and it breaks, they cannot get compensation from the network. However, they can seek compensation from people or organizations they trusted to supply a correctly functioning computer or communication service.''~\cite{shneiderman2000designing} Based on this definition, Shneiderman developed a set of guidelines for  developers of online services, such as e-commerce or e-services, which are underlined by two key principles. First, the organization providing the service should ensure trust both by providing evidence of past trustworthy performance and providing strong assurances of trust. Second, the organization should clarify responsibilities and obligations by providing full disclosure of terms, guarantees, and mechanisms for disputes. 

\brendan{This is quite mechanistic - I wonder what a biological living set of guidelines would look like...natural law? It'd be good to get Richman's view on this.}

Of course, when considering technology-mediated trust, it is also important to know that people have different perceptions of and attitudes towards technology, and so technology-mediated trust will be subjective~\cite{grabner2002role}. This is in line with the characteristics of trust in general by McKnight and Chervany which state that a person's own disposition to trust impacts trust formation.

\section{Trust of Software} 
In line with this, studies have shown that for software products, trust is based on both a trust of the creators of the software product and a trust of the software itself~\cite{sollner2016different,jackson2009direct}. Similarly, Siau and Wang argue that trust in technology is determined by three main factors: human characteristics, environment characteristics, and technology characteristics~\cite{siau2018building}. Users of software products assess the trustworthiness of software in different ways~\cite{jackson2009direct, yang2018unified, grodzinsky2011developing}. Yang et al. propose a software trust framework which considers software correctness, security, and reliability as measures of trustworthiness~\cite{yang2018unified}. Jackson equates trust to dependability of the software product to perform a particular task~\cite{jackson2009direct}. These definitions all relate to the intrinsic ability and internalized norms of the software. On the other hand, Wang et al. show that user feedback of software products is useful to determining levels of trust, which relates to the reputation of the software~\cite{wang2019updating}. Mercuri considered the view of transparency as it relates to trust of software, defining different ways that transparency can be achieved~\cite{mercuri2005trusting}. For example, through open sourced code, certifications, and assurances. Provenance, defined as ``metadata about the origin, context or history of data'', can also promote transparency~\cite{cheney2009provenance}. 

Another perspective comes from the field of human computer interaction where the relationship between trust and user interface design has been studied. Interface designers argue that the visual design of the interface forms the first impressions of trust~\cite{weinschenk2011100}. Rendell et al. found that inclusion of nature imagery on websites positively influenced users' perceptions of trust~\cite{rendell2021nature}. Xiling found that simple and well laid out interfaces promoted trust~\cite{xiling2005effective}. They also found that familiarity, being able to clearly relate the ``offline'' brand and experience to the online interface, was important for trust. Xiling found that usability was important for building trust~\cite{xiling2005effective}. Systems that were easy to use, consistent, and logically structured were more trusted. While researchers have investigated the use of particular colors in an interface and their relationship with user trust, no relationship was identified~\cite{hawlitschek2016colors}. 

\brendan{Are these superficial trust indicators? Trusting brands and interfaces is  the role of the dark arts' of brand/ advertising /sales etc...which is riding a very fine line at teh core of trust and building for example 'trust' in Veracity Lab. This is associated with a level of enquiry/ authenticity and vulnerability/ exposure to truth trust and...authenticity.}

\section{Trust of AI} 
With the rise of Artificial Intelligence (AI) to perform decision-making, it is also important to consider trust as it relates to these systems in particular. Glikson and Woolley find that the representation of an AI system (e.g., robot, humanoid, embedded) and the system's capabilities are important factors in developing trust~\cite{glikson2020human}.
Siau and Wang present a list of factors that are used for both building initial trust in AI systems and developing continuous trust in those systems~\cite{siau2018building}. They find that understanding how AI works (its transparency and explainability) and being able to trial the AI system before adopting it (trialability) are important for initial trust formation. In addition, the visual appearance of the AI (its representation), reviews of the AI system written by other users, and the users' perceptions of AI in general based on exposure to things like media coverage or Sci-fi books will also impact initial trust. 

\brendan{Q: Why are sci-fi books etc so dystopian? They nearly all indicate a lack of trust in our own work. Its a kinda double psychology twisted look at ourselves. }

While many AI systems operate as black boxes (there is no way to understand why decisions are being made), transparency, explainability, and interpretability are still seen as important for trust of AI systems~\cite{lipton2018mythos}.  Some research defines interpretable as understandable or transparent~\cite{lou2013accurate}. Others define interpretable as providing explanations for decisions. In these cases, the model may not be transparent, but some understandable reasons for decisions are provided by the black box AI model~\cite{lou2012intelligible}.  Thus, interpretability may be defined as both explainable or transparent.   Explanations are often proposed to improve trust in AI systems~\cite{lim2019these, sadeghi2021cases} and recent research shows that software users do want explanations when complex decisions are being made~\cite{chazette2020explainability}. 

For developing continuous trust in AI systems, Siau and Wang ~\cite{siau2018building} find that usability and reliability, collaboration and communication, sociability and bonding, security and privacy protection, interpretability, concerns about job replacement, and goal congruence are important factors. Of course, accuracy is also important. Yin et al. found that people considered both a model's stated accuracy and its observed accuracy in determining their trust of the model~\cite{yin2019understanding}. Siau and Wang say ``trust in AI
takes time to build, seconds to break, and forever to
repair once it is broken!''~\cite{siau2018building}

We have seen many examples where AI has gone wrong, and Winfield and Jirotka argue that ethical governance is critical to building
trust in AI~\cite{winfield2018ethical}. Through a  literature review of trust and AI, Lockey at al. identified five main challenges: 1)
transparency and explainability, 2) accuracy and
reliability, 3) automation resulting in job loss, 4) anthropomorphism (or including human-like characteristics) leading to over-estimation of the AI system, and
5) privacy concerns related to mass data extraction~\cite{lockey2021review}.

Another important factor related to trust in AI is fairness~\cite{vakkuri2020current}. While one might assume machines can make more fair decisions that are free from human bias, it is well known that AI systems actually amplify existing bias~\cite{chauhan2022role, ntoutsi2020bias}. Historical bias in training data can cause AI systems to learn this bias and make biased decisions.

Accountability is also important~\cite{vakkuri2020current} for trust in AI. This factor considers who will be held responsible for the decisions made by AI systems. There is currently not a clear answer to who should be held accountable. The Law Commission of England and Wales and the Scottish Law Commission recently proposed that self driving car users should not be held responsible for crashes and other driving offenses.\footnote{\href{https://www.forbes.com/sites/zacharysmith/2022/01/25/self-driving-car-users-shouldnt-be-held-responsible-for-crashes-uk-report-says}{https://www.forbes.com/sites/zacharysmith/2022/01/25/self-driving-car-users-shouldnt-be-held-responsible-for-crashes-uk-report-says}}. However, in the US, self driving car users are considered responsible. Research suggests more auditing of AI is needed to reduce corporate reputation damage and assure AI is legal, ethical, and safe~\cite{koshiyama2022algorithm}. 

Trust of AI also comes down to perceptions of how decisions are made. Machines make decisions which are rule-based and algorithmic~\cite{dietvorst2015algorithm}. Machines do not consider emotions in their decision-making nor can they learn in the same way as humans~\cite{cuzzolin2020knowing}. These differences can lead to algorithmic aversion, where people prefer human made decisions even if the decisions are inferior to those made by a machine~\cite{jussupow2020we}. 

Jussupow et al. defined four characteristics of algorithms that influence aversion: 1) algorithm agency which describes the level at which the algorithm behaves autonomously; 2) algorithm performance which considers the accuracy and failures of the algorithm; 3) perceived algorithm capabilities which describes the algorithm's perceived ability to perform the task; and 4) human involvement which relates to how much humans (but not the end user) are involved in training and using the algorithm~\cite{jussupow2020we}.

Recent research has found that people do not want AI to make moral decisions~\cite{bigman2018people}. Through a series of studies, Bigman and Gray found that people distrust AI to make moral decisions even when the outcome is favorable since ``machines can neither fully think nor feel''~\cite{bigman2018people}. 
They suggest that limiting AI to providing only advice and increasing the AI's perceived experience and expertise are ways to improve trust in AI for moral decisions~\cite{bigman2018people}. 

\begin{figure}
    \centering
    \includegraphics{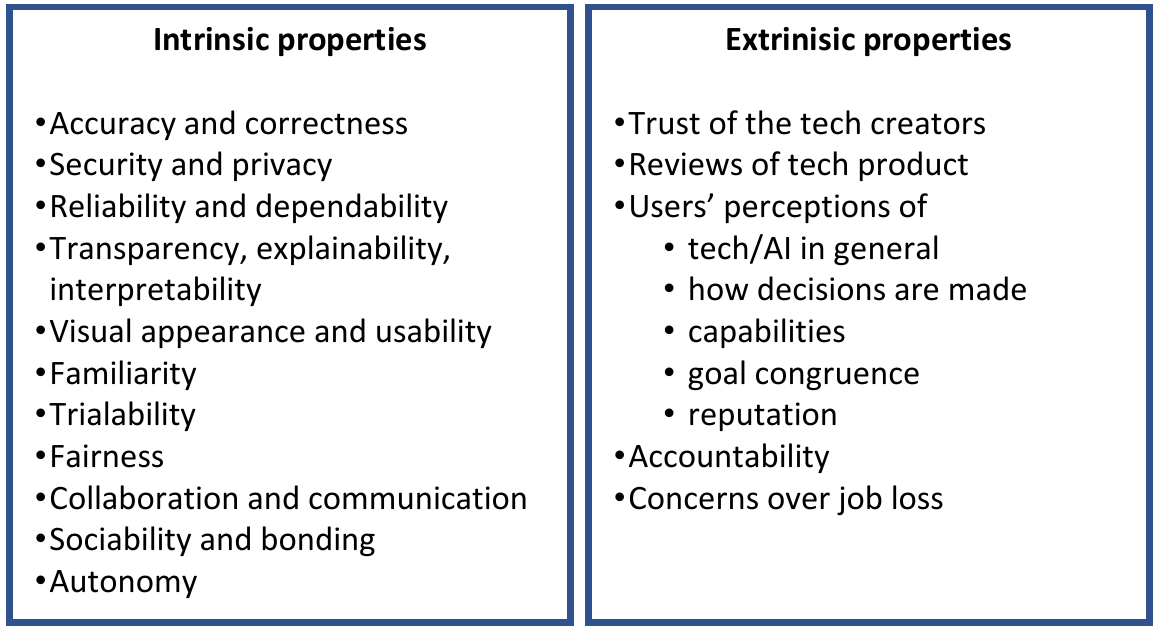}
    \caption{Factors that affect trust of digital technology}
    \label{fig:trust_tech}
\end{figure}

Figure~\ref{fig:trust_tech} summarizes the factors that influence trust for digital technologies, including software products and AI.

\section{Blockchain}
A discussion on trust of software is not complete without a mention of Blockchain. Blockchain has been nicknamed a ``trustless'' technology~\cite{vidan2019mine,hawlitschek2018limits}. It emerged due to a growing lack of trust in centralized systems which relied on trust of institutions (e.g. banks)~\cite{de2020blockchain}. The idea of a trustless technology goes back to the theory of Wang and Emurian that claims that trust is only needed  when there is some level or risk or uncertainty involved~\cite{wang2005overview}. Blockchain is a distributed, immutable ledger. This means transactions cannot be modified once they are written to the ledger and all participants have access to the shared ledger. Thus, ``users subject themselves to the authority of a technological system that they are confident is immutable, rather than to the authority of centralized institutions which are deemed untrustworthy.''~\cite{de2020blockchain}

De Filippi et al. prefer to label Blockchains as ``confidence machines'', since their underlying technology creates shared expectations and confidence in the correctness of its transactions.
However, while many Blockchains remove the need to trust a single organization, they still require ``distributed trust'' since there are often a large number of actors who require a low-level of trust~\cite{de2020blockchain}. There still needs to be trust that the data going into the Blockchain can be trusted since compromised data cannot be corrected. These actors must be trusted not to collude and cause collective harm. It should also be noted that not all Blockchains follow this same model and some (e.g. The Linux Foundation's Hyperledger and Amazon's QLDB) are maintained by organizations, which means these organizations will also still require trust. Using blockchain may lead to trade-offs between trust and other concerns like energy consumption and sustainability~\cite{sedlmeir2020energy}. These tradeoffs could in turn compromise trust because benevolence and integrity can be adversely affected if a Blockchain is perceived to be non-sustainable. 

There does not appear to be literature on trust violations and repair in relation to blockchain technology.

\brendan{If I could conclude by where I started on our overall commitment to Mātauranga Maori and data/ computer/ trust etc By doing so it could take us on quite a remarkable journey of exploring the topic of this paper and subsequent interrelated topics like fairness, accountability obligation etc 
Thanks}

\bibliographystyle{ieeetr}
\bibliography{main}

\end{document}